\UseRawInputEncoding
\documentclass[superscriptaddress,twocolumn,amsmath,longbibliography,amssymb]{revtex4-2}
\usepackage{graphicx}
\usepackage{siunitx}
\usepackage{hyperref}
\usepackage[capitalize]{cleveref}
\usepackage{bm}
\usepackage{amsfonts}
\usepackage{booktabs}
\usepackage{array}
\usepackage{tabularx}

\begin{document}  

\title{On the conditional equivalence of phase retrieval algorithms}

\author{J.~Schr\"oder}
\email{jakob.schroeder@physik.uni-muenchen.de}
\affiliation{Fakult\"at f\"ur Physik, Ludwig-Maximilian-Universit\"at M\"unchen, Am Coulombwall 1, 85748 Garching, Germany}
\affiliation{Munich Center for Machine Learning (MCML)}

\author{A.~D\"opp}
\affiliation{Fakult\"at f\"ur Physik, Ludwig-Maximilian-Universit\"at M\"unchen, Am Coulombwall 1, 85748 Garching, Germany}
\affiliation{Munich Center for Machine Learning (MCML)}
\affiliation{Max Planck Institut f\"ur Quantenoptik, Hans-Kopfermann-Strasse 1, Garching 85748, Germany}

\begin{abstract}
Phase retrieval---recovering a complex-valued field from intensity measurements---is typically solved using variants of the Gerchberg--Saxton (GS) algorithm, understood as alternating projections between measurement planes. Meanwhile, modern computational imaging increasingly relies on gradient-based optimization and automatic differentiation. Here we show that these two approaches are mathematically identical: the GS magnitude replacement step is exactly a unit gradient descent step on an amplitude least-squares loss. This equivalence enables seamless integration of classical phase retrieval with differentiable physics pipelines. We further identify two complementary probabilistic interpretations of this equivalence: globally, the amplitude loss is the negative log-likelihood under Gaussian amplitude noise; locally, each projection step arises as a Bayesian update with the propagated field as prior. The local view provides qualitative guidance for relaxation in iterative phase retrieval.
\end{abstract}

\maketitle

\section{Introduction}

Phase retrieval is the problem of recovering a complex-valued signal from magnitude-only measurements and arises throughout science and engineering \cite{shechtmanPhase2014}. 
It underpins coherent diffraction imaging at X-ray and electron sources \cite{miaoCrystallography2015, marinelliSingleShot2013}, astronomical wavefront sensing \cite{fienupPhaseretrieval1993}, and the design of diffractive optical elements \cite{vorndranBroadband2015}. 
In laser and accelerator physics specifically, phase retrieval is central to spatio-spectral characterization of high-power laser pulses \cite{smartsevSimple2024} and to beam diagnostics based on coherent radiation spectra \cite{bakkalitaheriElectron2016,
schmidtLongitudinal2018a}. 
In all these settings phase often encodes the quantity of interest, yet detectors measure intensity.

For over fifty years, the Gerchberg-Saxton (GS) algorithm \cite{gerchbergHolography1972} has been the workhorse of practical phase retrieval. The algorithm is elegantly simple: propagate a field estimate between measurement planes, replacing the magnitude with measured data while preserving the phase, and iterate until convergence. This ``alternating projections'' picture provides clear physical intuition.

The limitations of basic GS, like stagnation at local minima, slow convergence and sensitivity to noise, have motivated numerous extensions. The hybrid input-output (HIO) method \cite{fienupPhase1982} introduces a relaxation parameter to escape stagnation. Relaxed averaged alternating reflections (RAAR) \cite{lukeRelaxed2004} provides improved convergence guarantees. Regularized variants incorporate prior knowledge such as support constraints, smoothness, or sparsity \cite{sidorenko_sparsity-based_2015, chang_phase_2016}. Yet these modifications share a common feature: their parameters like relaxation weights or regularization strengths are typically tuned for each application, often justified by empirical success rather than first principle derivations.

Modern computational imaging has increasingly adopted gradient-based optimization for phase retrieval, enabled by differentiable physics simulators and automatic differentiation frameworks \cite{kandel_using_2019, wong_phase_2021, du_adorym_2021}. In this framework, phase retrieval is posed as minimizing a loss that quantifies the mismatch between predicted and measured data. Modifications---regularization, learned priors, additional constraints---enter transparently as terms in the objective, and gradients are computed via backpropagation through the optical forward model.

In this letter, we show that these approaches are mathematically identical. The GS magnitude replacement step is exactly a unit gradient descent step on an amplitude least-squares loss. This equivalence has immediate practical value: GS can be implemented via automatic differentiation frameworks, enabling seamless integration with neural networks, learned priors, and hybrid physics-ML pipelines without sacrificing the physical interpretability of the classical algorithm. For instance, phase retrieval pipelines for laser diagnostics \cite{smartsevSimple2024} that embed GS as an inner loop could be made end-to-end differentiable, enabling joint optimization of retrieval parameters alongside upstream processing steps.

We call this equivalence conditional as it only holds for free-space propagation, but not for systems with absorption, magnification, or sampling changes. We would like to mention that Fienup already observed in his seminal 1982 paper \cite{fienupPhase1982} that for the single-intensity problem error reduction is equivalent to steepest descent on the amplitude-squared error, and that for the two-intensity problem steepest descent with respect to the phase approximates the GS update up to a position-dependent weighting factor and a small-phase linearization. The present work establishes the exact unit-step identity for the two-intensity case by optimizing instead over the complex field with an amplitude loss, removing both approximations. Since Fienup's work predates backpropagation by four years \cite{rumelhart1986learning}, the conceptual bridge to modern differentiable programming was not available at the time.

A complementary line of work, initiated by Levi and Stark \cite{leviImage1984} and systematized by Bauschke, Combettes and Luke \cite{bauschkePhase2002}, recasts phase-retrieval algorithms as operator compositions in the framework of convex feasibility. This projection-operator view is complementary to ours: it describes the structural correspondence between classical phase retrieval and convex-optimization algorithms, while the gradient view we develop here describes the update rule in a form directly compatible with automatic differentiation and composable with additional loss terms.

Separately, the Wirtinger-flow family \cite{candesPhase2015a, weiConjugate2017} applies gradient-based optimization to an intensity-based loss under random measurement models (Gaussian sampling or coded diffraction patterns), relying on spectral initialization for its convergence guarantees. This setting differs from the structured unitary measurement geometry of classical GS considered here, and the amplitude loss we employ is well-behaved from random initialization.

Furthermore, we show that the same amplitude loss arises independently as the negative log-likelihood under a Gaussian noise model on measured amplitudes. This statistical interpretation is not required for the GS--gradient descent equivalence, but it is exploitable: the probabilistic framework makes explicit the assumptions implicit in the amplitude loss, and provides a principled route to extensions. Priors on the source field become regularization terms; noise models with known variance suggest specific relaxation parameters. Crucially, because these extensions simply modify the loss function, they remain fully differentiable---the two perspectives reinforce each other.

\section{Gerchberg--Saxton as Gradient Descent}

Consider a complex-valued source field $u \in \mathbb{C}^N$ to be recovered from amplitude measurements at two planes. Let $A$ denote the linear propagation operator between planes (typically a discrete Fourier transform for far-field measurements), and let $a_1, a_2 \in \mathbb{R}^N_{>0}$ denote the measured amplitudes at planes 1 and 2 respectively. That is, we observe $a_1$ and $a_2$ and seek a field $u$ satisfying $|u| = a_1$ and $|Au| = a_2$, where we assume $A$ is unitary, so $A^{-1} = A^*$.

The Gerchberg--Saxton algorithm constructs such a field iteratively. Starting from an initial guess $u^{(0)}$ (typically $u^{(0)} = a e^{i\phi}$ for random phase $\phi$), each iteration proceeds as follows:
\begin{align}
    v^{(t)} &= A u^{(t)}, \label{eq:gs_forward}\\
    \tilde{v}^{(t)} &= a_2  \frac{v^{(t)}}{|v^{(t)}|}, \label{eq:gs_replace2}\\
    \tilde{u}^{(t)} &= A^* \tilde{v}^{(t)}, \label{eq:gs_back}\\
    u^{(t+1)} &= a_1 \frac{\tilde{u}^{(t)}}{|\tilde{u}^{(t)}|}. \label{eq:gs_replace1}
\end{align}
Steps \eqref{eq:gs_replace2} and \eqref{eq:gs_replace1} are the characteristic operations: replace the magnitude with measured data while preserving the phase. The algorithm alternates between enforcing the amplitude constraint at plane 2 and at plane 1, with propagation steps in between.

This magnitude replacement has a geometric interpretation. Define the amplitude constraint set $\mathcal{M}_a = \{u \in \mathbb{C}^N : |u| = a\}$---a product of circles in complex space, one per pixel. The operation $u \mapsto a (u/|u|)$ is precisely the Euclidean projection onto $\mathcal{M}_a$:
\begin{equation}
    P_a(u) \;=\; a \frac{u}{|u|} \;=\; \underset{z \,:\, |z| = a}{\mathrm{argmin}} \|z - u\|^2.
    \label{eq:projection}
\end{equation}

The GS algorithm is thus ``alternating projections''---project onto the constraint set at plane 2, propagate back, project onto the constraint set at plane 1, propagate forward, and repeat.

Now consider the amplitude loss
\begin{equation}
    \mathcal{L}_a(u) = \frac{1}{2} \| |u| - a \|^2,
    \label{eq:amp_loss}
\end{equation}
which measures the squared distance between the amplitude of $u$ and the target $a$. The gradient with respect to the complex field is 

\begin{equation}
    \nabla_u \mathcal{L}_a = (|u| - a) \frac{u}{|u|} = u - a \frac{u}{|u|} = u - P_a(u).
    \label{eq:amp_gradient}
\end{equation}
A gradient descent step with unit step size therefore gives
\begin{equation}
    u - \nabla_u \mathcal{L}_a = u - (u - P_a(u)) = P_a(u).
    \label{eq:main_identity}
\end{equation}
Consequently, the projection onto the amplitude constraint and the unit gradient step on the amplitude loss are the same operation.

The full GS iteration follows by applying this observation at each plane. Define the loss at plane 2 as a function of the source field:
\begin{equation}
    \mathcal{L}_{a_2}(u) = \frac{1}{2} \| |Au| - a_2 \|^2.
\end{equation}
By the chain rule, $\nabla_u \mathcal{L}_{a_2} = A^* (Au - P_{a_2}(Au))$. A unit gradient step yields
\begin{equation}
    u - \nabla_u \mathcal{L}_{a_2} = u - A^* Au + A^* P_{a_2}(Au) = A^* P_{a_2}(Au),
\end{equation}
using unitarity $A^* A = I$. This is exactly the GS half-iteration \eqref{eq:gs_forward}--\eqref{eq:gs_back}: propagate forward, replace magnitude, propagate back. The same reasoning applies to the constraint at plane 1. The standard GS algorithm is thus equivalent to alternating unit gradient descent steps on the losses $\mathcal{L}_{a_1} = \frac{1}{2}\||u| - a_1\|^2$ and $\mathcal{L}_{a_2} = \frac{1}{2}\||Au| - a_2\|^2$.

Why amplitude rather than intensity? One might instead consider the intensity loss $\mathcal{L}^{\mathrm{int}}(u) = \frac{1}{2}\| |u|^2 - I \|^2$, whose gradient is $\nabla_u \mathcal{L}^{\mathrm{int}} = 2(|u|^2 - I) u$. This has a fatal flaw: the gradient vanishes whenever $u = 0$, even if the target intensity $I$ is nonzero. Dark pixels in the estimate remain stuck regardless of the measured data. The amplitude loss avoids this pathology, which explains why GS---implicitly using amplitude coordinates---succeeds where naive intensity-based gradient descent fails.

The equivalence established here has immediate practical consequences. First, GS can be implemented by defining the amplitude loss and calling a standard gradient descent step; no explicit projection code is required, and the algorithm integrates directly with automatic differentiation frameworks. Second, the loss-based view enables hybrid pipelines: additional terms (regularizers, learned priors, constraints from other measurements) enter as additive contributions to the objective, and gradients propagate through the entire system. Third, convergence results for gradient descent apply directly to GS, providing a unified analytical framework.

Precisely these additional terms can be motivated further. The amplitude loss \eqref{eq:amp_loss} was introduced here on geometric grounds. In the next section, we show that the same loss arises from a probabilistic formulation, which clarifies its implicit assumptions and suggests principled extensions.

\section{A Probabilistic Perspective}

\subsection{Global inference and regularization}

Here we show that the same amplitude loss arises independently from a probabilistic formulation. This coincidence is not merely aesthetic. The statistical framework makes explicit the assumptions hidden in the amplitude loss, and provides a principled route to extensions. Priors on the source field, noise models with known variance, and regularization all enter systematically, with parameters derived from the physics rather than tuned by hand. Because these extensions amount to modifying the loss function, they remain fully compatible with the gradient-based view of Section~II.

We frame phase retrieval as inference in a generative model. The source field $u$ produces fields at the two measurement planes via deterministic propagation: $u$ at plane 1, $Au$ at plane 2. The measured amplitudes $a_1$ and $a_2$ are noisy observations of $|u|$ and $|Au|$ respectively. The inference task is to recover $u$ given $a_1$ and $a_2$.

The simplest noise model assumes Gaussian fluctuations on the measured amplitudes:
\begin{equation}
    p(a_1 \mid u) \propto \exp\left( -\frac{1}{2\sigma_1^2} \| |u| - a_1 \|^2 \right),
    \label{eq:likelihood1}
\end{equation}
and similarly
\begin{equation}
    p(a_2 \mid u) \propto \exp\left( -\frac{1}{2\sigma_2^2} \| |Au| - a_2 \|^2 \right).
    \label{eq:likelihood2}
\end{equation}
The measurements at the two planes are conditionally independent given $u$, so the joint likelihood is
\begin{equation}
    p(a_1, a_2 \mid u) = p(a_1 \mid u) \, p(a_2 \mid u).
\end{equation}
Maximum likelihood estimation seeks the field $u$ maximizing this expression, or equivalently minimizing the negative log-likelihood:
\begin{equation}
    \hat{u} = \underset{u}{\mathrm{argmin}} \left[ \frac{1}{2\sigma_1^2} \| |u| - a_1 \|^2 + \frac{1}{2\sigma_2^2} \| |Au| - a_2 \|^2 \right].
    \label{eq:mle}
\end{equation}
In the case of equal noise levels ($\sigma_1 = \sigma_2 = \sigma$), the prefactors cancel and we recover exactly the sum of amplitude losses from Section~II.

The amplitude loss is the negative log-likelihood under Gaussian amplitude noise. This is a modeling choice, not a physical derivation. Poisson statistics are often more appropriate for photon-counting detectors. We discuss this in the following subsection.

The GS algorithm performs this optimization by alternation: a gradient step on $\mathcal{L}_{a_2}$ (enforce amplitude at plane 2), then a gradient step on $\mathcal{L}_{a_1}$ (enforce amplitude at plane 1), and repeat. This is a form of coordinate descent, minimizing the joint objective by cycling through its components. That such alternating optimization converges to the same solution as joint gradient descent on $\mathcal{L}_{a_2}$ is an established result precisely in this MLE picture \cite{nealView1998}. The alternating structure is a computational choice, not a fundamental feature of the probabilistic model.

Prior knowledge about the source field enters through Bayes' theorem. With prior $p(u)$, maximum a posteriori (MAP) estimation replaces \eqref{eq:mle} with
\begin{equation}
    \hat{u} = \underset{u}{\mathrm{argmin}} \left[ \mathcal{L}_{a_1}(u) + \mathcal{L}_{a_2}(u) - \log p(u) \right].
    \label{eq:map}
\end{equation}
The term $-\log p(u)$ acts as a regularizer: support constraints, smoothness penalties, and sparsity-inducing norms all arise from specific prior choices. The regularization strength is then determined by the ratio of measurement noise variance to prior variance, suggesting it should be set based on signal-to-noise considerations rather than tuned ad hoc.

This global view---MLE or MAP on the joint objective---explains regularization but does not address \emph{relaxation}: modifications to individual projection steps, as in HIO or RAAR, that improve convergence without adding explicit priors. For this we turn to a local picture of inference at each measurement plane.

\subsection{Local inference and gain-controlled projections}

Consider a single pixel at one measurement plane; we suppress the plane index for clarity. Let $v \in \mathbb{C}$ denote the propagated prediction at that pixel, and let $u \in \mathbb{C}$ denote the latent field value we wish to estimate. The propagation step provides prior information about $u$. To expose the implicit assumptions of the GS update, we ask: under what local noise model does magnitude replacement arise as the optimal estimator? The simplest such model assigns a circular complex Gaussian prior centered at $v$:

\begin{equation}
    p(u \mid v) \propto \exp\!\left( -\frac{|u - v|^2}{2\sigma_p^2} \right),
    \label{eq:local_prior}
\end{equation}
where $\sigma_p^2$ quantifies uncertainty in the propagated estimate. It pairs it with a Gaussian likelihood on the measured amplitude:

\begin{equation}
    p(a \mid u) \propto \exp\!\left( -\frac{(|u| - a)^2}{2\sigma_a^2} \right).
    \label{eq:local_likelihood}
\end{equation}

The local posterior is $p(u \mid a, v) \propto p(u \mid v)\, p(a \mid u)$. We do not claim that this model is physically correct -- intensity would for instance be governed by Poisson statistics -- but rather that it is the model under which the standard GS step is the optimal estimator for the $\sigma_a\rightarrow 0$ limit we discuss below.

This posterior couples amplitude and phase, but asymmetrically. The likelihood is phase-invariant, so the posterior mode inherits the phase of $v$---explaining why GS ``carries the phase forward'' as a conditional mode rather than an ad hoc choice. As for the amplitude replacement, the MAP of this model takes the form
\begin{equation}
    u^{+} = v + \alpha\,(a - |v|)\,\frac{v}{|v|} = (1 - \alpha)\,v + \alpha\,a\,\frac{v}{|v|},
    \label{eq:gain_update}
\end{equation}
with gain
\begin{equation}
    \alpha = \frac{\sigma_p^2}{\sigma_p^2 + \sigma_a^2}.
    \label{eq:gain}
\end{equation}
Equation~\eqref{eq:gain_update} interpolates between retaining the prediction ($\alpha = 0$) and full magnitude replacement ($\alpha = 1$). It admits two equivalent readings: as a Bayesian MAP update with gain $\alpha$, and as a gradient descent step on the amplitude loss with learning rate $\alpha$. 

The standard GS projection is the limiting case $\alpha \to 1$, which arises when the measurement is noise-free ($\sigma_a^2 \to 0$) or when the prediction is uninformative ($\sigma_p^2 \gg \sigma_a^2$). The latter regime describes early iterations from random initialization, where the propagated field carries little information and enforcing the measured amplitude fully is appropriate. As the reconstruction improves, however, the prediction becomes informative and $\sigma_p^2$ should shrink. Continuing with $\alpha \equiv 1$ then amounts to assuming insufficient confidence in an increasingly informative prediction, which can cause the oscillatory or stagnant behavior observed in practice.

This local Bayesian picture provides a statistical interpretation of relaxation. The gain~\eqref{eq:gain} suggests that relaxation parameters should reflect the ratio of prediction uncertainty to measurement noise, and that stronger relaxation (smaller $\alpha$) becomes appropriate as the reconstruction converges. While $\sigma_p^2$ is not directly computable without expensive covariance tracking, this interpretation offers qualitative guidance: relaxation should strengthen over iterations, and its magnitude should be informed by the measurement signal-to-noise ratio. The single-variable model captures the noise-driven aspect of relaxation while relaxation methods such as HIO and RAAR additionally address nonconvex-stagnation escape via reflection-based updates, which is structurally outside the model considered here and is rigorously accounted for by the operator-splitting framework of Bauschke et al.~\cite{bauschkePhase2002}.

\section{Numerical Demonstration}

\subsection{Equivalence and loss choice}

We verify the identity \eqref{eq:main_identity} numerically. A $128 \times 128$ test image (camera man) provides the amplitude $a_1^{\mathrm{meas}}$. We attach a random phase to create a ground-truth field, propagate via unitary DFT to obtain $a_2^{\mathrm{meas}}$, and recover the field starting from a zero initial phase.

We compare:
\begin{enumerate}
    \item \textbf{GS:} Explicit magnitude replacement per \cref{eq:gs_forward}--\eqref{eq:gs_replace1}
    \item \textbf{GD:} Gradient descent alternating on $\mathcal{L}_{a_1}$ and $\mathcal{L}_{a_2}$ with $\alpha = 1$, gradients computed via automatic differentiation
\end{enumerate}

\begin{figure}
    \raggedright
    \noindent\includegraphics{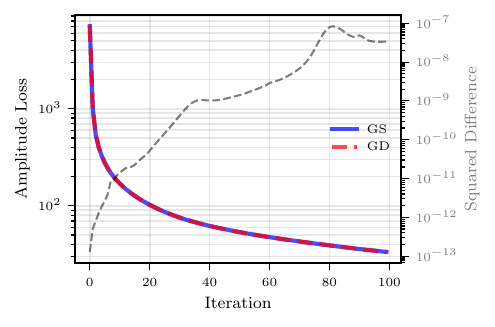}
    \caption{Amplitude loss versus iteration for GS (alternating projections) and GD (autodiff with unit step size). The right y-axis shows the mean squared difference between the field estimates of both methods at each iteration.}
    \label{fig:equivalence}
\end{figure}

\Cref{fig:equivalence} shows near-identical convergence for both methods. Residual inconsistencies between the respective field estimates are assumed to stem from numerical differences in the respective functions. For example, backpropagation through the automatic differentiation of a forward FFT could slightly differ from the inverse FFT function.  

\begin{figure}
    \raggedright
    \noindent\includegraphics{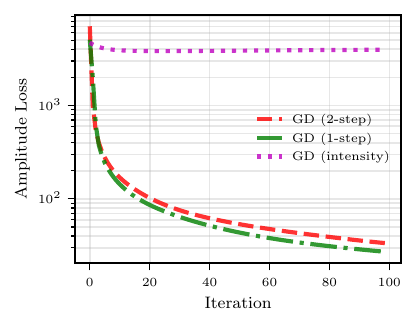}
    \caption{Amplitude loss versus iteration for three different implementations of GD, with gradients being calculated for: the losses on the amplitude in both planes in an alternating fashion with 2 steps (the same case as in Figure \ref{fig:equivalence} which was equivalent to GS), the sum of losses in both planes in 1 step, the losses on the intensity in both planes in 2 steps.}
    \label{fig:comparison}
\end{figure}

The equivalence also enables exploration of alternative optimization strategies. Figure \ref{fig:comparison} compares GS to joint gradient descent on the sum of amplitude losses ("1-step"), which converges faster by avoiding the coordinate-descent structure of alternating projections. The intensity loss, by contrast, stagnates due to vanishing gradients at low-amplitude pixels. This pathology of the intensity loss explains why Wirtinger-flow methods \cite{candesPhase2015a, weiConjugate2017} rely on spectral initialization to land inside a basin of attraction.

\subsection{Scheduled learning rate under noise}

\begin{figure}
    \raggedright
    \noindent\includegraphics{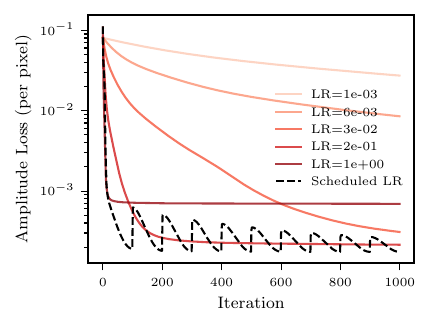}
    \caption{Amplitude loss per pixel versus iteration at five fixed learning rates and under a cosine-annealed schedule with warm restarts. Both measured amplitudes are corrupted by $5\%$ Gaussian noise. The unit-step case corresponds to classical GS.}
    \label{fig:lr_schedule}
\end{figure}

The local Bayesian picture predicts that $\alpha$ in \eqref{eq:gain_update} should shrink as the reconstruction improves, while its specific functional form is not determined by the model. Figure~\ref{fig:lr_schedule} tests this qualitative prediction on the same cameraman task with $5\%$ Gaussian amplitude noise, comparing fixed learning rates to a cosine-annealed schedule with decaying restarts as one generic decreasing schedule from the optimization literature. Unit-step GS (corresponding to a learning rate of 1.0) plateaus before the optimum as a result of its implicit overconfidence in the noisy data, whereas some fixed sub-unit learning rates reach close to it.  The schedule reaches consistently below all fixed choices, already before the first restart. 
The restarts are retained as a precaution against landscape stagnation but appear unnecessary on this problem. HIO and RAAR address such stagnation through reflection-based updates whose operator-splitting interpretation~\cite{bauschkePhase2002} does not arise naturally from the local Bayesian model.

\subsection{Phase reconstruction with an overcomplete dictionary}

\begin{figure}
    \raggedright
    \noindent\includegraphics{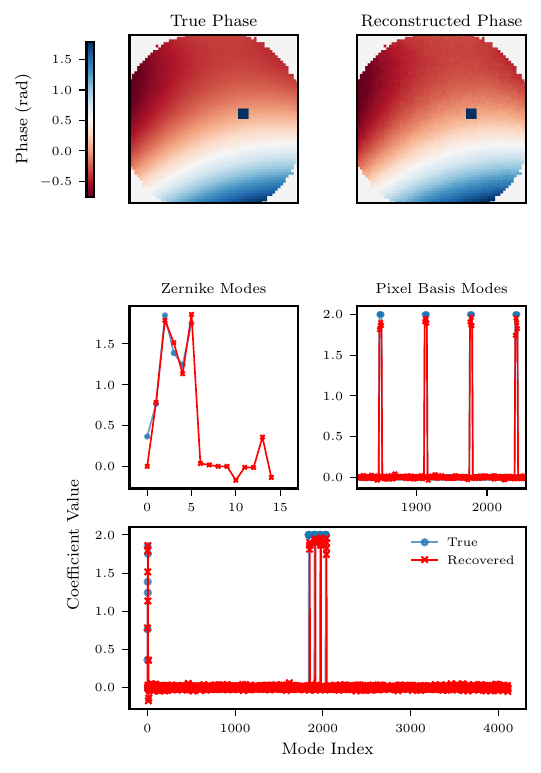}
    \caption{Phase reconstruction with an overcomplete dictionary of Zernike polynomials and pixel-basis atoms. Top: ground truth and reconstructed phase. Middle and bottom: true (blue) and recovered (red) coefficients in the Zernike and pixel blocks. Source data uses a tophat amplitude profile corrupted by $2\%$ Gaussian noise. Test data uses fewer Zernike modes than the reconstruction dictionary contains, probing the overcomplete fit.}
    \label{fig:modal}
\end{figure}

To illustrate composition with a differentiable upstream model of the source field, we parameterize the phase as an expansion over an overcomplete dictionary combining low-order Zernike polynomials with a pixel basis~\cite{howard_sparse_2025} and minimize the amplitude loss with an added $\ell_1$ penalty over the dictionary parameters using Adam~\cite{kingmaAdamMethodStochastic2017}. The test phase mixes a few Zernike modes with a localized hot-pixel deviation with measurements carrying $2\%$ amplitude noise. Figure~\ref{fig:modal} shows that recovery is accurate in both dictionary blocks, correctly identifying the smooth and localized components that neither basis alone would capture.

Because the amplitude loss is exactly the objective that underpins the classical GS algorithm, its empirical track record transfers directly to parameterized settings where only the representation of the input has changed. Further upstream components or additional loss terms compose with the same objective without structural change to the optimization. Hybrid schemes that combine GS with modal expansions through non-gradient updates\cite{moulanier_fast_2023} require separate architectural work, and thus, implementation effort, to accommodate each extension.

\section{Conclusion}

We have shown that the Gerchberg--Saxton algorithm and gradient descent on an amplitude loss are mathematically identical: the magnitude replacement step is exactly a unit gradient step, producing the same iterates to sufficient precision. This is not a new algorithm but a new understanding of an existing one, with practical consequences for how phase retrieval integrates with modern computational tools.

The equivalence enables GS to be implemented via automatic differentiation without explicit projection code. More significantly, it allows classical phase retrieval to participate fully in differentiable physics pipelines: additional constraints, learned priors, and hybrid physics-ML architectures enter as terms in a loss function, with gradients propagating through the entire system. The physical interpretability of GS is preserved while gaining the flexibility of gradient-based optimization.

We further showed that the amplitude loss arises independently as the negative log-likelihood under Gaussian amplitude noise. This statistical interpretation makes explicit what GS implicitly assumes, and provides a principled framework for extensions. Priors become regularizers; unequal noise levels suggest weighted objectives. The geometric and probabilistic views converge on the same algorithm, reinforcing confidence in both.

A complementary perspective emerges from considering inference at a single measurement plane. Modeling the propagated field as a Gaussian prior and the amplitude measurement as a likelihood yields a gain-controlled update that interpolates between retaining the prediction and enforcing the measurement. 
The standard GS projection is the limiting case of infinite measurement confidence or uninformative prediction and is appropriate for early iterations from random initialization, but increasingly suboptimal as the reconstruction improves. While the relevant uncertainty is not directly computable without expensive covariance tracking, the interpretation offers qualitative guidance. This local Bayesian picture captures the noise-driven aspect of relaxation; methods such as HIO and RAAR additionally target nonconvex-stagnation escape through reflection-based updates, whose rigorous account is the operator-splitting framework.~\cite{bauschkePhase2002}.

Several limitations deserve mention. The equivalence requires unitary propagation. Non-unitary propagation (absorption, magnification, sampling changes, partial coherence) breaks the unit-step identity, since information is lost in the forward step and thus the round-trip operator is no longer the identity. The differentiable-pipeline framing nevertheless extends to any differentiable forward model. Prior knowledge encoded as regularization (as in Eq.~\eqref{eq:map}) then becomes essential to constrain the unresolved modes.
The Gaussian amplitude noise model is a convenient idealization---Poisson statistics are often more appropriate for photon-counting detectors and would yield a different loss function. However this idealization is implicitly in accordance with the original GS algorithm.
The connection between the gain parameter and existing relaxation schemes provides interpretation rather than closed-form derivation. A fully probabilistic treatment of reflection-based methods such as HIO and RAAR would likely require a joint latent-variable model, of which the local model considered here can be understood as a limiting case, and is left to future work.
Finally, while joint optimization on both planes (Figure~\ref{fig:comparison}) converges faster than the alternating structure of GS, it requires computing gradients through both propagation steps simultaneously, presenting a memory-compute tradeoff for large-scale problems.

\begin{acknowledgments}
This work was supported by the Independent Junior Research Group "Characterization and control of high-intensity laser pulses for particle acceleration", DFG Project No.~453619281. J.S. thanks the International Max Planck Research School for Advanced Photon Science (IMPRS-APS) for support.
\end{acknowledgments}

\bibliography{references_gs}

\end{document}